# ERUPTION OF PROMINENCES TRIGGERED BY CORONAL RAIN IN THE SOLAR ATMOSPHERE OBSERVED BY SDO/AIA AND STEREO/EUVI

**Z. M. Vashalomidze,[1] T. V. Zaqarashvili,[1,2,3] V. D. Kukhianidze,[1] and G. T. Ramishvili[1]**

*The triggering process for coronal mass ejections (CME) in the solar atmosphere is not fully understood. We use observations from different spacecraft at several wavelengths to detect an instability process for a prominence/filament with subsequent eruption of CME. Time series of spectral lines at 304, 171, 193, and 211 Å have been obtained with the SDO spacecraft, and at 304, 171, 195, and 284 Å with the STEREO spacecraft. A prominence/filament system was observed during November 8-23, 2011, at different angles by SDO, STEREO_A, and STEREO_B. The observations show that a giant tornado began to develop near the base of the prominence at 20:00 UT on November 20, which later caused the appearance of droplets of coronal rain (at 16:00 UT, November 21) which fell downward from the main mass of the prominence. The coronal rain continued until 20:20 UT, November 22, and caused an instability of the prominence, after which a CME took place 22:30 UT on November 22. We assume that the loss of mass owing to coronal rain may lead to instability of prominences and their subsequent eruption. Observations of coronal rain falling from the main part of the mass of prominences could be used for predicting space weather.*

Keywords: *coronal rain: solar corona: solar atmosphere*

[1] E. K. Kharadze Abastumani Astrophysical Observatory, Ilia State University, Tbilisi, Georgia, e-mail: zurab.vashalomidze.1@iliauni.edu.ge, vaso@iliauni.edu.ge
[2] Space Research Institute, Austrian Academy of Sciences, Schmiedlstrasse 6, 8042 Graz, Austria, e-mail: teimuraz.zaqarashvili@uni-graz.at
[3] IGAM-Kanzelhohe Observatory, Institute of Physics, University of Graz, Universitatsplatz 5, 8010 Graz, Austria

## 1. Introduction

Solar prominences/filaments are cold, dense plasma structures in the hot solar corona which are sustained by a coronal magnetic field that compensates the force of gravity [1,2]. Sometimes prominences lose stability and erupt in the form of CME, which ultimately influences space weather conditions on the earth [3,4]. The triggering of CME is probably related to some process leading to instability in the solar atmosphere [5] that is not currently known [6-8]. Observations show that the dynamic behavior of prominences/filaments is usually controlled by the coronal magnetic field [9-11]. One of the important processes for initiation of CME is the tornado instability which sometimes shows up in the solar limb in the form of dark vertical structures and can be seen clearly in pictures of the sun in hot coronal spectral lines [12-16]. A tornado usually forms alongside the legs of solar filaments and prominences [12,15,17,13,18]. Recent observations by SDO (the Solar Dynamics Observatory) show that almost all the tornados associated with prominences become unstable and half of them lead to CME; thus a tornado could be used as an effective instrument for prediction of space weather [19].

Another important process in the solar atmosphere is coronal rain, a sequence of cold, dense plasma droplets (blobs) falling along solar coronal loops toward their base. Coronal rain is probably caused by a thermal instability [20,21]; it is also related to solar prominences where the cold droplets separate from the main mass and fall downward toward the photosphere. Recent observations from the AIA (Atmospheric Imaging Assembly) on board the SDO [23,24] have shown that the bulk of the mass of filaments is not static but is maintained by condensation with a high estimated velocity compared to comparable drainage through numerous vertically downward flows [22]. These flows in the form of cold dense bunches show up as coronal rain falling from a height of 20-40 Mm. The droplet velocity has a narrow Gaussian distribution with a mean of 30 km/s, while the downward moving accelerating distribution had an exponential fall with an average of 46 m/s$^2$. It was assumed that the thermal instability leading to formation of the coronal rain proceeds mainly through catastrophic cooling when the radiative losses locally exceed the heating [25-27]. Numerical simulation also shows that catastrophic cooling owing to thermal instability may be the reason for formation of the cold condensation and, therefore, of the coronal rain [28-30].

When the plasma of a prominence begins to fall in the form of coronal rain, its mass obviously decreases. Thus, the upward directed Lorentz force ultimately overcomes the gravitational force and this process can lead to instability of the prominence. In this paper we study the formation, dynamics, and instability of solar prominences using observations from SDO/AIA and STEREO/EUVI, and the role of coronal rain in the triggering of CME.

## 2. Observation and analysis of data

Observations were made with the SDO and STEREO (Solar-Terrestrial Relationships Observatory) on November 8-23, 2011. AIA/SDO observes the sun in channels at many wavelengths, providing images with high spatial resolution of 0.6 arcsec/pixel with a 12 s cycle [23,24]. We used four extreme ultraviolet narrow band filters, 304, 171, 193, and 211 Å with corresponding temperatures of $10^{4.7}$, $10^{5.8}$, $10^{6.1}$, and $10^{6.3}$ K. The set of instruments "for study of coronal and heliospheric coupling of the sun and earth" (SECCHI) on board the space ship STEREO is a set of five telescopes which received images of the sun in extreme ultraviolet bands (304, 171, 195, 284 Å) using the EUVI (Extreme Ultraviolet Imager) [31-33]. We used only 304 and 195, as well as 171 Å if it showed up in the data sets. SECCHI/EUVI provides images with a high spatial resolution of 1.6 arcsec/pixel of the entire solar disk (image size 2048x2048 pixels) [31-33]. The SDO and STEREO satellites observe the sun at different angles, which makes it possible to follow the structure and dynamics of a prominence in detail. The separation angle between SDO and STEREO_B (STEREO_A) during our observations was -104° (+106°) relative to SDO. At the same time the line at the western edge of the SDO image (the western limb) lies +14° inside STEREO_B, while the line at the eastern edge of SDO is at +16° inside STEREO_A (Fig. 1). In order to identify specific points on the sun in both STEREO and SDO, we used the SSW (Solar Software) procedure from the STEREO base STEREO wcs_convert_diff_rot, which can be used to determine the position of specific points in images from both of the spacecraft.

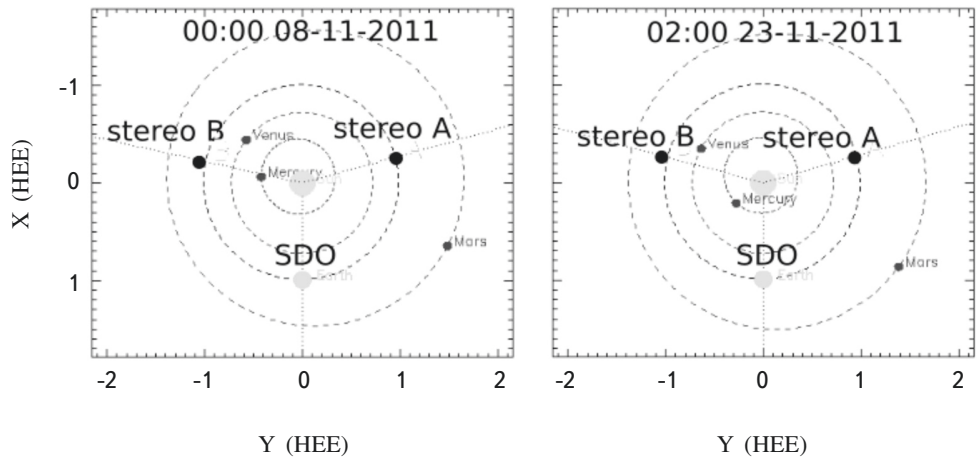

Fig. 1. The position of SDO, STEREO_A, and STEREO_B during the observation period (November 2011). The line at the western edge of the SDO image (western limb) lies at +14° inside STEREO_B and the line at the eastern edge of SDO, at +16° inside STEREO_A.

Observations of the structure and dynamics of a prominence were made during November 2011. A prominence/filament was observed in the STEREO_B pictures up to November 14. In addition, on SDO it showed up on November 8 and was noticed until November 23. When the prominence first showed up at the western limb in the SDO images, it was noticed on a disk near the western limb in the STEREO_B images. Figure 2 shows the system of prominences/filaments on November 10 (top frames) and November 14 (bottom frames). The left frames show composite (different wavelengths in a single image) images from STEREO_B and the right frames, composite images from SDO/AIA.

The prominence intersected the entire solar disk for the next few days. On November 7 it showed up in images from the STEREO_A spacecraft, while it was still visible in the eastern part of the solar disk on SDO/AIA, and on November 20, it reached the eastern limb. Figure 3 shows a system of prominences on SDO and STEREO_A on November 18 (top frames) and 20 (bottom frames). The left and right frames show analogous composite images as in Fig. 2. From 20:16 UT (November 20) to 12:14 UT (November 21) from the lower regions near the support of the prominence a tornado began to rise (Fig. 4). The tornado can be seen more distinctly in the STEREO_A/EUVI images as a dark vertical structure (right column).

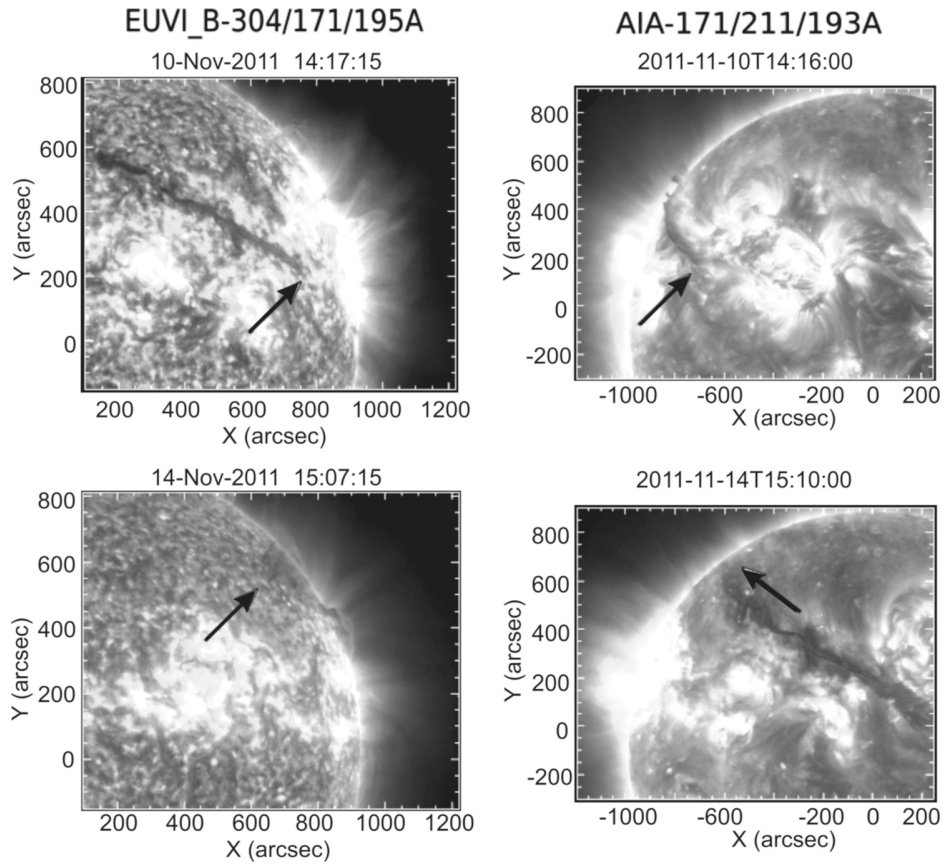

Fig. 2. The evolution of the prominence/filament over November 10-14. The left column shows composite images in EUVI, made up of three spectrum lines, 171, 195, and 304 Å from STEREO_B at 14:17 UT on November 10 (top frame) and 15:07 UT on November 14 (bottom frame). The right column shows composite images made up of three spectrum lines, 171, 193, and 211 Å from SDO/AIA at 14:16 UT on November 10 (top frame) and 15:10 UT on November 14 (bottom frame). The black arrows in the top frames indicate the right edge of the prominence viewing at different angles. The arrows in the bottom frames indicate the left edge of the prominence seen by the two spacecraft.

When the tornado reached its maximum height, droplets of rain began to fall from the main mass of the filament. The rain began at 16:00 UT on November 21. Figure 5 shows the coronal trajectories of the rain (along the coronal loops) as seen by both instruments. The plasma flowed out for almost thirty hours, which means that a substantial fraction of the mass of the prominence was depleted as rain. The white arrows indicate the location near the surface where the droplets of coronal rain fell.

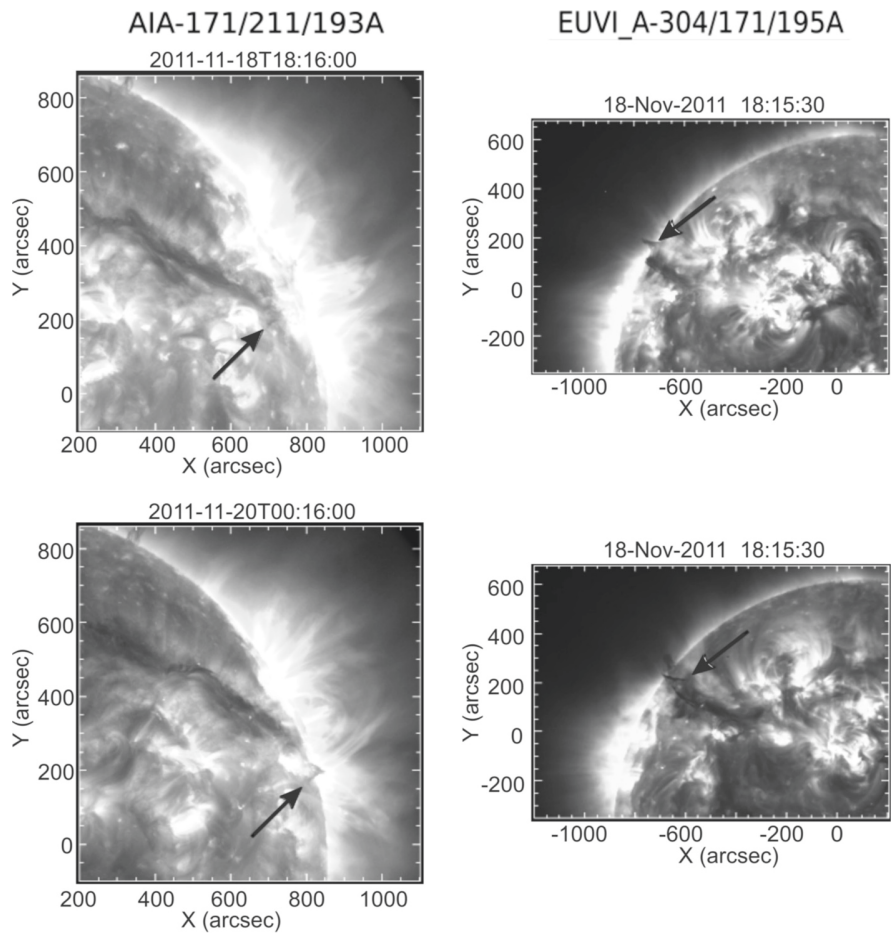

Fig. 3. Evolution of the prominence/filament during November 18-20. The left column shows composite images from SDO/AIA at 18:16 UT, November 18 (top frame) and at 16:00 UT, November 20 (bottom frame). The right column shows composite images in EUVI/STEREO_A at 18:15 UT, November 18 (top frame) and at 15:31 UT, November 20 (bottom frame). The black arrows indicate the location of the tornado near the base of the filament.

The comparative brightness of these regions shows that either the droplets of rain are heated in the corona until the time when they fall or they heat the chromosphere plasma during a collision.

Figure 6 illustrates the dynamics of a single rain droplet over 30 min. These frames show the region bounded by the dark rectangle in the upper left frame of Fig. 5. The black arrows indicate the position of the droplet at a certain (specific) time. The droplet moved about 70 Mm over the 30 min with an average velocity of 40 km/s.

After 28 hours of coronal rain (mass loss), the prominence became unstable. Figure 7 shows the temporal dynamics of the prominence over 20:00-24:00 UT on November 22 as seen from the side of SDO and STEREO_A. The instability of the prominence sets in at about 20:20 UT (top frames), when the material of which it was composed moved upward. After an hour (middle panels) the prominence has already moved substantially upward (this can be seen especially well in the left frames). After about three hours (about 23:30 UT) the prominence is no longer seen in the lines of AIA and EUVI (bottom frames), since it has been ejected as a CME.

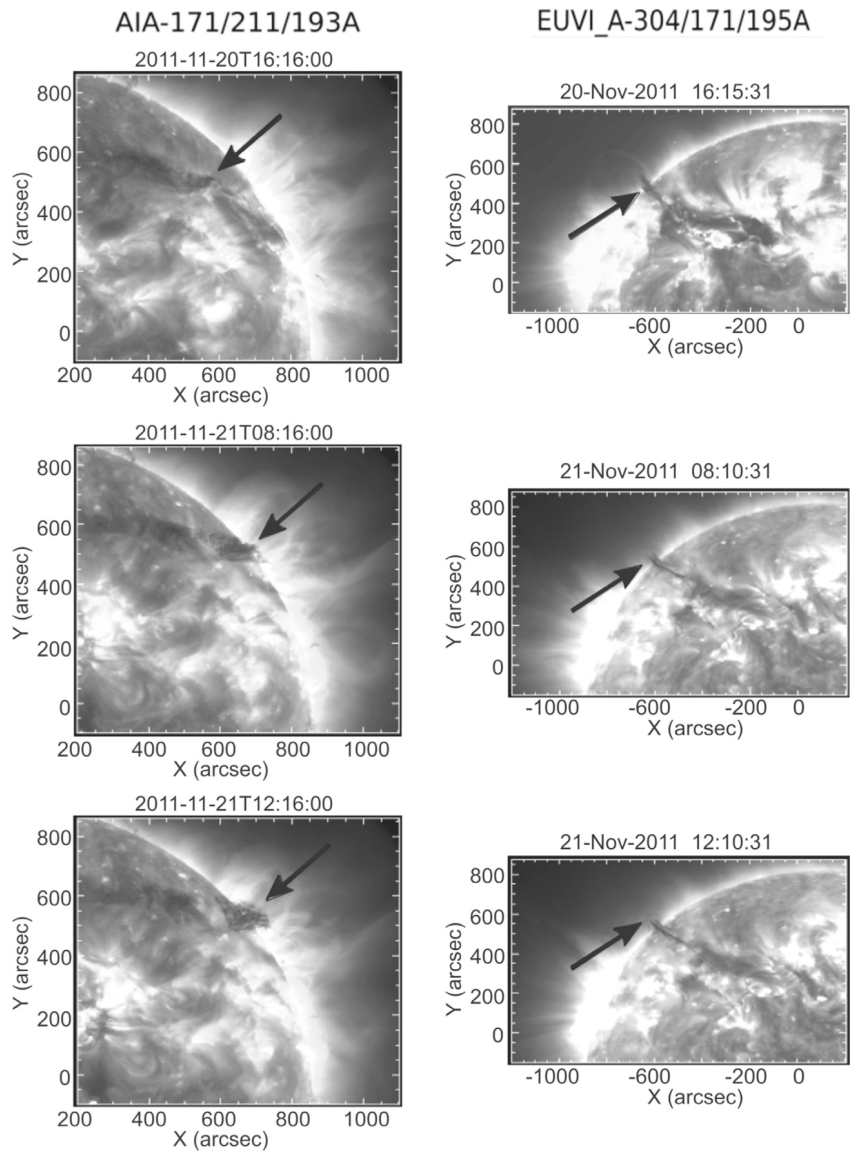

Fig. 4. The rise of a tornado at the base of the prominence on November 20-21. The left column shows composite images from SDO/AIA; at 16:16 UT on November 20 (top frame), 08:16 UT on November 21 (middle frame), and 12:16 UT on November 21 (bottom frame). The right column shows composite images from EUVI/STEREO_A: at 16:15 UT on November 20 (top frame), 08:10 UT on November 21 (middle frame), and 12:10 UT on November 21 (bottom frame). The black arrows show the evolution of the tornado.

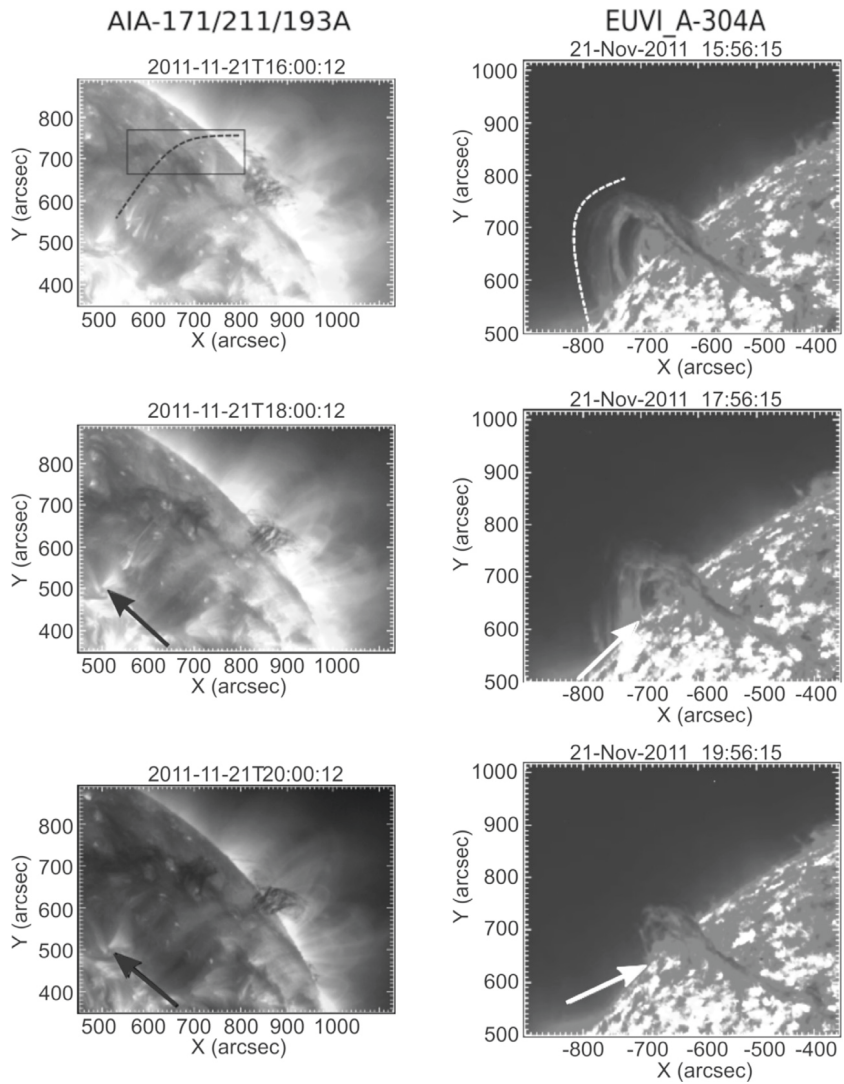

Fig. 5. Coronal rain on November 21 from SDO and STEREO_A. The left column shows composite images from SDO/AIA. The top, middle, and bottom frames correspond to 16:00, 18:00. and 20:00 UT, respectively, on November 21. The black rectangle indicates the region where the droplets fall, as shown in Fig. 6. The dashed curve is the trajectory of the falling droplets, and the black arrows indicate the places where they fall. The right column shows the 304 Å channel in EUVI/STEREO_A. The top, middle, and bottom frames correspond to 15:56, 17:56. and 19:56 UT, respectively, on November 21. The white arrows indicate the same points in the STEREO images as marked by the black arrows in the SDO images.

When the filament vanished in the coronal lines it obviously appeared in white light images from the STEREO-A coronagraphs (COR1 and COR2) as a CME. Figure 8 shows the temporal dynamics of the CME in the outer corona as observed from STEREO-A. COR1 observes in visible light at distances of 1.5-4 solar radii from the center, and COR2, from 2-15 solar radii. It is clear that the CME showed up in the COR1 images at 22:30 UT, i.e., an hour after it rose from the lower corona. Thus, the average velocity at which the CME rose was about 200 km/s.

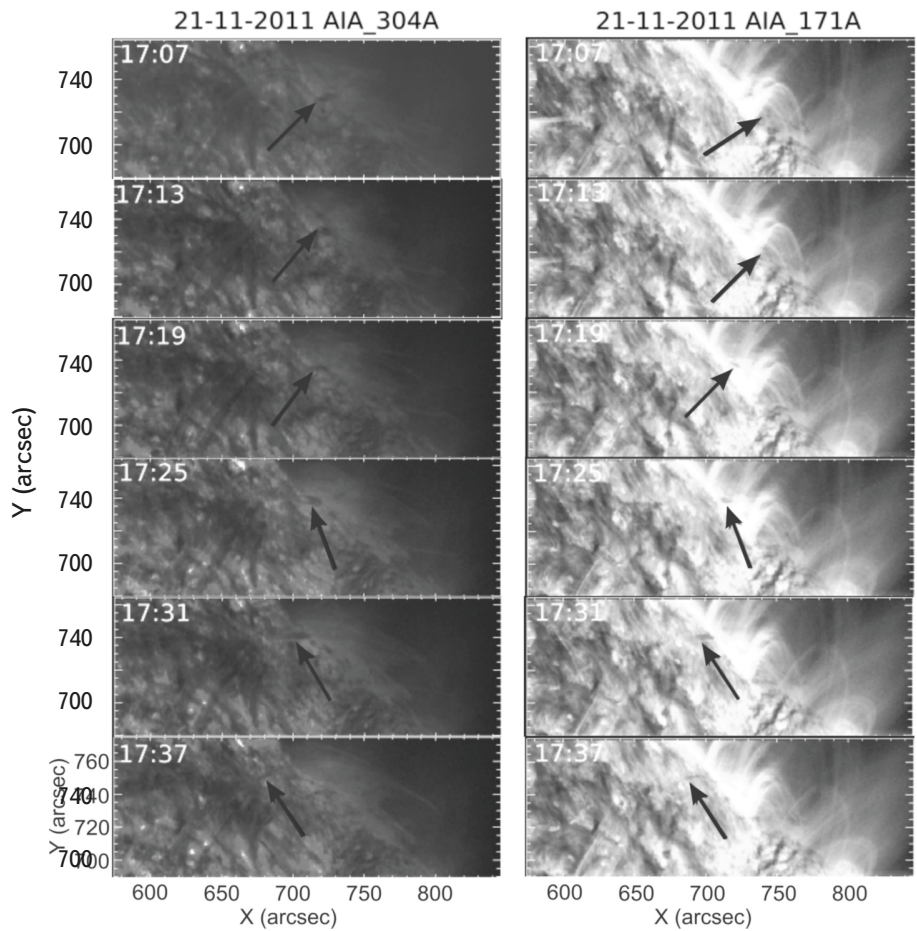

Fig. 6. The motion of a single droplet of rain in the corona in the region of the black rectangle indicated in the upper left frame of Fig. 5. The dynamics of the coronal rain droplet is shown over 30 min with an interval of 6 min between successive frames (from top to bottom). The left and right columns show images in the 304 and 171 Å lines, respectively, of SDO/AIA. The black arrows indicate the location of the coronal rain droplet at the successive times.

The temporal dynamics of the prominence showed us that the instability that led to the CME began after an almost 30 hour outflow of droplets of coronal rain from the main body (mass) of the prominence. The following scenario is proposed for explaining the development of the instability of the prominence. Filaments are almost 100 times denser that the surrounding coronal plasma so they should fall rapidly to the surface under the influence of gravity. But the gravitational force is balanced by the magnetic field which is probably stretched because of the large mass of the prominence. If the coronal rain continuously removes part of the mass from the filament, then the gravitational force is ultimately weakened and this should lead to an upward rise of the remaining part of the prominence because of the excess Lorentz force. This may lead to magnetic reconnection or a similar process, which produces a global instability of the prominence and, therefore, the CME.

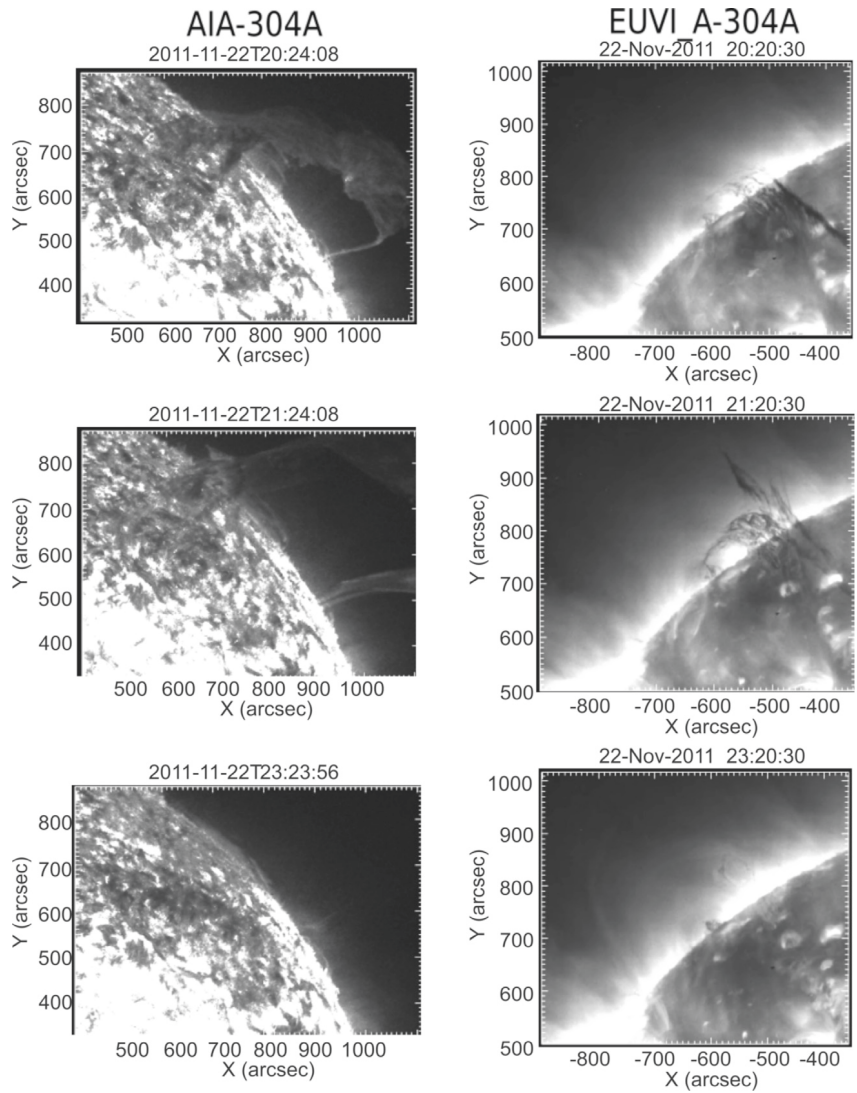

Fig. 7. Destabilization (loss of equilibrium) and eruption (ejection) of a prominence after coronal rain. The left column shows three different 304 Å images in SDO/AIA. The top, middle, and bottom frames correspond to 20:24, 21:24, and 23:24 UT, respectively, on November 22. The right column shows three different images in the 193 Å channel of STEREO_A. The top, middle, and bottom frames correspond to 20:20, 21:20, and 23:20 UT on November 22.

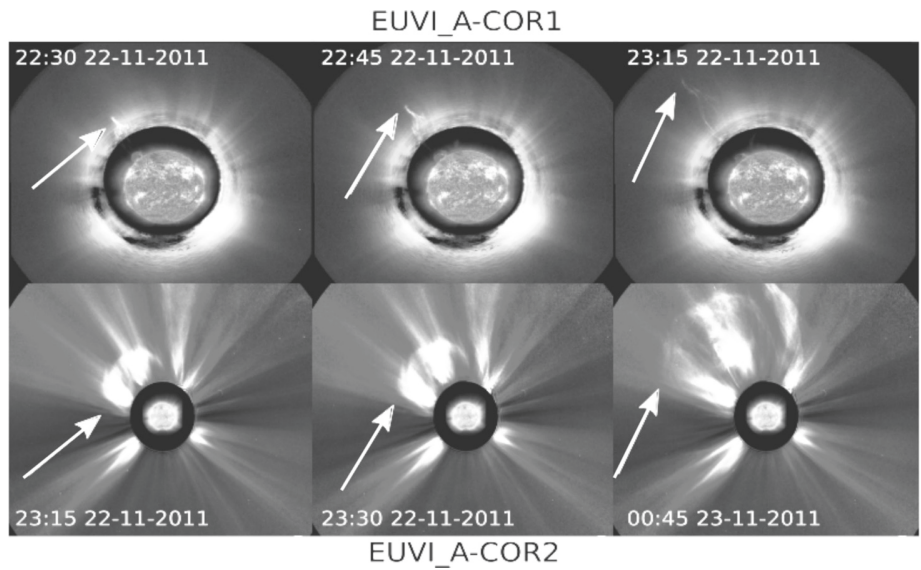

Fig. 8. A CME eruption as seen from STEREO_A. The top frames show three successive images of the CME in COR1 STEREO_A. COR1 displays an image in white light at distances of 1.5-4 solar radii from the sun's center. The sun in EUVI-A_304Å is shown in the central part of each image. The white arrows indicate the location of the CME. The CME appeared in the window of COR1 at about 22:30 on November 22 and left it at 23:15 (top right frame). The bottom frames show the time evolution of the CME in COR2 STEREO_A. COR2 displays an image in white light at distances of 2-15 solar radii from the sun's center. The sun in EUVI-A_195Å is shown in the central part of each image.

## 3. Conclusions

A solar prominence was observed by several space craft at different wavelengths from 00:00 UT on November 8 to 00:45 UT on November 23, 2011. We used observational data from the AIA/SDO, EUVI/STEREO_A and EUVI/STEREO_B satellites.

SDO, STEREO_A and STEREO_B observed the sun from three different angles, so we could track the detailed dynamics of the prominence/filament with high spatial resolution. During the observed time interval the prominence intersected the entire solar disk in images from SDO. We observed the formation of a tornado at the base of the prominence which ultimately began to grow at roughly 20:00 UT on November 20. When the tornado reached the height of the prominence, coronal rain began to fall from its base region at about 16:00 UT on November 21. After about 30 hours of rain, the system became unstable at 20:20 UT and erupted as a CME at 22:30 UT on November 22. We assume that the mass loss owing to the coronal rain led to instability of the prominence and provoked the CME. If future observations show that coronal rain is a widespread process for initiation of CME, then it could be used for prediction of space weather.

This work was supported by the Shota Rustaveli National Science Foundation (SRNSF) [PhDF2016_147], grant DI-2016-52, and grant 217146.